\documentstyle[12pt]{article}

\def\a{\alpha }

\begin{document}

\begin{titlepage}

\begin{center}
{\Large\bf Polarized Parton Densities in the Nucleon}
\end{center}
\vskip 2cm

\begin{center}
{\bf Elliot Leader}\\
{\it Department of Physics\\
Birkbeck College, University of London\\
Malet Street, London WC1E 7HX, England\\
E-mail: e.leader@physics.bbk.ac.uk}\\
\vskip 0.5cm
{\bf Aleksander V. Sidorov}\\
{\it Bogoliubov Theoretical Laboratory\\
 Joint Institute for Nuclear Research\\
141980 Dubna, Russia\\
 E-mail: sidorov@thsun1.jinr.dubna.su}
\vskip 0.5cm
{\bf Dimiter B. Stamenov \\
{\it Institute for Nuclear Research and Nuclear Energy\\
Bulgarian Academy of Sciences\\
Blvd. Tsarigradsko Chaussee 72, Sofia 1784, Bulgaria\\
E-mail:stamenov@inrne.acad.bg }}
\end{center}

\vskip 0.3cm
\begin{abstract}
We present a critical assessment of what can be learnt from the
present data on inclusive polarized DIS. We examine critically
some of the simplifying assumptions made in recent analyses and
study in detail the question of the determination of the gluon, 
strange sea and valence quark polarized densities.

We have also carried out a new NLO QCD analysis of the world data.
We find an excellent fit to the data and present our results
for the polarized parton densities.
\end{abstract}
\vskip 0.5 cm

\end{titlepage}

\newpage

\setcounter{page}{1}

{\bf 1. Introduction.}
\vskip 4mm

Deep inelastic scattering (DIS) of leptons on nucleons has remained the
prime source of our understanding of the internal partonic
structure of the nucleon and one of the key areas for the testing
of perturbative QCD. Decades of experiments on unpolarized
targets have led to a rather precise determination of the unpolarized
parton densities. Spurred on by the famous EMC experiment \cite{EMC} 
at CERN in 1988, there has been a huge
growth of interest in polarized DIS experiments which yield more
refined information about the partonic structure. Many
experiments have been carried out at SLAC \cite{SLACnQ2}-\cite{E154}
and CERN \cite{SMCd93_95}-\cite{newSMCpQ2} on proton, deuterium
and  $^{3}He$ targets, and there are major programmes under way at SLAC 
(E155), DESY (HERMES \cite{HERMES}) and CERN (COMPASS).\\

In addition to the unpolarized structure functions $F_1(x,Q^2)$ and 
$F_2(x,Q^2)$ there are two independent spin structure functions   
$~g_1(x,Q^2)~$ and $~g_2(x,Q^2)~$ and their unambiguous
determination requires measurement of both the longitudinal
asymmetry $A_{\parallel}$ and the transverse asymmetry $A_{\bot}$
obtained with a target polarized parallel or perpendicular to the 
lepton beam direction, respectively.
In the recent years there has been a great improvement in the quality
of the data on the structure function
$~g_1(x,Q^2)~$, obtained from measurements using a longitudinally
polarized target, and a big extension in the kinematic range $x$ and 
$Q^2$ covered. Moreover it has become possible to present data in
bins of $~(x, Q^2)~$ rather than simply averaged over $Q^2$ at each 
$x$. The spin dependent structure function $~g_2(x,Q^2)~$ has now also
been extracted \cite {SLACnQ2, finSLACpd, g2} from the data although 
with limited statistical  
precision compared to the $g_1$ determination. 

The data at very small $x$ has taught us that extrapolation of the 
measured values to $~x=0~$ is a subtle matter, so that the moments 
of the structure functions should not be considered as genuinely 
experimental quantities. \\


Experiments on unpolarized DIS provide information on the
unpolarized quark densities $q(x,Q^2)$ and  gluon density $G(x,Q^2)$
inside a nucleon. Measurements of $~g_1(x,Q^2)~$ give us more 
detailed information,
namely the number densities of quarks $~q(x,Q^2)_{\pm}~$ and gluons 
$~G(x,Q^2)_{\pm}~$ whose helicity is respectively along or opposite 
to the helicity of the parent nucleon. The usual densities are
\begin{equation}
q(x,Q^2) = q_{+}(x,Q^2) + q_{-}(x,Q^2)~,~~~~  
G(x,Q^2) = G_{+}(x,Q^2) + G_{-}(x,Q^2) 
\label{unpolpa}
\end{equation}
and the new information is then contained in the polarized
structure function $~g_1(x,Q^2)~$ which is expressed in terms of
the {\it polarized} parton densities 
\begin{equation}
\Delta q(x,Q^2) = q_{+}(x,Q^2) - q_{-}(x,Q^2)~,~~~~
\Delta G(x,Q^2) = G_{+}(x,Q^2) - G_{-}(x,Q^2)~. 
\label{polpa}
\end{equation}

Several theoretical analyses \cite{GSt} - \cite{Bour}
based on the NLO perturbative QCD calculations \cite{nlocor} have 
sought to pin down the polarized parton densities.
Each of them utilized the different data sets available at the
time the analyses were performed. Only in the analyses 
\cite {Alta, MStr, DeRoeck} is essentially all the present data
used, the exception being the very recent final E143 results 
\cite{finSLACpd}). And none have completely used 
the information contained in the more detailed binning of the data in 
$~(x,Q^2)$. Moreover, there are differences in the assumptions
to aid the analysis, differences in the choice of the
renormalization scheme and differences in the form of the input parton
densities and the value $Q^2_0$ at which they are determined. And
finally, there are still significant disagreements about the results. 
To quote one example, Altarelli et al. \cite {Alta} have 
obtained a significant polarized gluon density $~\Delta
G(x,Q^2)~$, whereas Bourrely et al. \cite {Bour} claim that a
perfectly acceptable fit can be obtained with $~\Delta G(x,Q^2) =
0~$.\\

Our aim in this paper is twofold:

i) We discuss what can be learnt from the data, in theory and in
practice, and examine the role played by the various assumptions
used in the theoretical analyses.

ii)We carry out a new study of the world polarized data. 
In addition to the data used in our previous analysis \cite {LSiSt}
the more accurate SLAC/E154 neutron data \cite {E154}, the new
more precise SMC proton data \cite {newSMCpQ2} which do not indicate 
a rise of
$g_1$ at small $x$, the HERMES data \cite {HERMES} and the final
data results \cite {finSLACpd} of the E143 Collaboration  at SLAC
are now included. In trying to extract as well as possible the polarized 
parton densities we pay special attention to the observed scaling
violations in the $A_1$ data and use all the information contained
in the more detailed binning of them in $(x,Q^2)$.
As in our previous work we use the following
parametrization for the input polarized parton densities:
\begin{equation}
\Delta q(x,Q^2_0)=f(x)q(x,Q^2_0)~,
\label{inputpa}
\end{equation}
in which we now utilize the {\it new} MRST set of unpolarized densities  
$~q(x,Q^2_0)~$ \cite{MRST}. These parton densities account for the
new, more precise H1 and ZEUS deep inelastic scattering data, for
the re-analysis of the CCFR neutrino data, for the inclusive prompt
photon and large $E_T$ jet production in proton-proton collisions 
and for the charge asymmetry in Drell-Yan reactions. The MRST
analysis leads to a value $~\alpha_s(M^2_{z}) = 0.1175~$ in
excellent agreement with the world average $~\alpha_s(M^2_{z}) =
0.118 \pm 0.005~$ \cite{alsaver}.\\

{\bf 2. Ambiguities and subtleties in determining the polarized
parton densities}
\vskip 4mm

There are several difficulties, specific to the polarized case,
which make it much harder than in the unpolarized case to obtain
reliable and unambiguous information about the polarized densities.\\

{\it 2.1. What can be deduced in principle}\\

In the unpolarized case the separation of the contributions from
partons of different flavours relies heavily upon the existence
of both charged current neutrino and neutral current
electromagnetic data. At present, and for some years to come, the
information from polarized inelastic measurements will be limited
to neutral current data. This raises the interesting question as
to what one can hope to learn, both in theory and in practice.

We have available data on $~g_1^{(p)}(x,Q^2)~$ and $~g_1^{(n)}(x,Q^2)~$
(or $~g_1^{(d)}(x,Q^2)~$) structure functions expressed as linear 
combinations of either the individual parton densities 
\begin{equation}
\Delta u +\Delta\bar{u},~~~~\Delta d + \Delta\bar{d},~~~~ \Delta s
+\Delta\bar{s}
\label{indpart}
\end{equation}
and $~\Delta G~$ or, equivalently, the $SU(3)$ flavour combinations
\begin{eqnarray}
\nonumber
\Delta q_3 = (\Delta u +\Delta\bar{u}) - (\Delta d + \Delta\bar{d})~,\\
\nonumber 
\Delta q_8 = (\Delta u +\Delta\bar{u}) + (\Delta d + \Delta\bar{d})
- 2(\Delta s+\Delta\bar{s})~,\\ 
\Delta \Sigma = (\Delta u +\Delta\bar{u})+(\Delta d + \Delta\bar{d})
+(\Delta s+\Delta\bar{s}) 
\label{flavpart} 
\end{eqnarray}
and $~\Delta G$.
(Note that $\Delta q_8$ is defined in such a way that its first
moment is $\sqrt {3}$ times the expectation value of the eighth
component of the Cabibbo axial vector $SU(3)$ current.)

We are trying therefore to obtain information about four
functions of $x$ and $Q^2$ on the basis of experimental data on
the two independent functions $~g_1^{(p)}(x,Q^2)~$ and 
$~g_1^{(n)}(x,Q^2)$. It is simpler to discuss the situation in
terms of the contributions $~\Delta q_{3},~\Delta q_{8},
~\Delta \Sigma~$ and $~\Delta G$.

We have
\begin{equation}
g_1^{p(n)}(x,Q^2)={1\over 2}({2\over 9})\{\delta C_{NS}\otimes 
[\pm {3\over 4}\Delta q_3 + {1\over 4}\Delta q_8] + 
\delta C_{S}\otimes \Delta \Sigma + 
\delta C_{G}\otimes \Delta G\}~,
\label{g1pn}
\end{equation}
where $\delta C_{NS},~\delta C_{S}$ and $\delta C_{G}$ are the 
non-singlet, singlet
and gluon Wilson coefficient functions, respectively. In (\ref
{g1pn}) $\otimes$ denotes convolution with respect to $x$ and
2/9 is the average value of the charge squared $<e^2>$ when the 
number of flavours $N_f=3$. Note that taking into account more
than 3 active flavours does not change the main conclusions in 
this and the next section.\\

Let us firstly suppose that we have perfect data for a range of
$x$ and $Q^2$, and that we try to determine the functions
$~\Delta q_{3},~\Delta q_{8},~\Delta \Sigma~$ and $~\Delta G$.
Then $\Delta q_3$ is determined uniquely and trivially since
\begin{equation}
g_1^{p}(x,Q^2)- g_1^{n}(x,Q^2) =
{1\over 6}\delta C_{NS}\otimes \Delta q_3 
\label{g1ns}
\end{equation}

We are then left with
\begin{equation}
g_1^{p}(x,Q^2) + g_1^{n}(x,Q^2) =
{2\over 9}[{1\over 4}\delta C_{NS}\otimes \Delta q_8 
+\delta C_{S}\otimes \Delta \Sigma + \delta C_{G}
\otimes \Delta G]~.
\label{g1s}
\end{equation}

It is the difference in the $Q^2$-evolution of the three terms on
the RHS of (\ref {g1s}) that enables them to be determined
separately. Indeed, by studying the first and higher derivatives 
of the LHS of (\ref {g1s}) with respect to $Q^2$ at $Q^2=Q^2_0$, 
and using the
evolution equations, one can prove that $\Delta q_{8},~\Delta 
\Sigma~$ and $~\Delta G$ are uniquely determined at $Q^2=Q^2_0$.

It is immediately clear, given the limited range of $Q^2$
available and the fact that the data are {\it not} perfect and
have errors, that the separation of $\Delta q_{8},~\Delta 
\Sigma~$ and $~\Delta G$ from each other will not be very
clearcut. Nonetheless, {\it in principle}, the data fix
$\Delta q_3,~\Delta q_{8},~\Delta \Sigma~$ and $~\Delta G$
or, equivalently, via (\ref{indpart}), $\Delta u +\Delta\bar{u},
~\Delta d + \Delta\bar{d},~ \Delta s+\Delta\bar{s}\equiv 
2\Delta\bar{s}$ 
and $\Delta G$. But any hope of a successful analysis will depend
upon finding simple enough parametrizations of these quantities at
$Q^2=Q^2_0$.\\

{\it 2.2. Valence and sea}\\

It is clear from the above that the inclusive (electromagnetic
current) data give no
information about the valence parts $~\Delta q_{v}~$ of the quark
densities. It is of interest, however, to know the 
$~\Delta q_{v}~$ and $~\Delta\bar{q}~$ or equivalently, $~\Delta
q~$ and $~\Delta\bar{q}~$, since they play a distinctive role in
other types of experiment, e.g. polarized semi-inclusive DIS,
polarized Drell-Yan reactions etc. Indeed, an attempt has been
made to extract the polarized valence densities from the
semi-inclusive data \cite{semiincl}, but the quality of the
present data precludes an accurate determination of these
densities. It is therefore important to make a combined analysis
\cite{deFlorian} of both the semi-inclusive and inclusive DIS data. 

Further, for the unpolarized densities simple parametrizations are
normally given for the valence and sea quark densities. One
reason for this is that one expects $q_v$ and $\bar{q}$ to have
simple, but very different, behaviours as $~x\to 0$, and this
feature is lost when dealing with $q$ or $q+\bar {q}$.

There is some point, therefore, in wanting to deal directly with
the valence and sea densities and to this end it has been common
practice to make some model assumptions about the polarized sea
\cite{GSt, Vog} and \cite{E154th}- \cite{Bour}, which then allows a 
determination of the valence parts.
For example, the apparently innocuous assumption of a flavour
independent polarized sea 
\begin{equation}
\Delta\bar{s}=\Delta\bar{u}=\Delta\bar{d}
\label{SU3}
\end{equation}
implies that $~\Delta u_{v}~$ and $~\Delta d_{v}~$ are determined, 
since the data fix $~\Delta\bar{s},~\Delta u_{v}+2\Delta\bar{u}~$ and 
$~\Delta d_{v}+2\Delta\bar{d}$.

We believe it is important to study the consequences of
assumptions like Eq. (\ref{SU3}). Consider therefore the family of
assumptions at $Q^2=Q^2_0$ 
\begin{equation}
\Delta\bar{u}=\Delta\bar{d}=\lambda \Delta\bar{s}~,
\label{SU3br}
\end{equation}
where $\lambda$ is a parameter.

Given that the data fix $~\Delta q_{3,8},~\Delta \Sigma~$ and
$~\Delta G~$ and that
\begin{equation}
\Delta\bar{s}={1\over 6}(\Delta \Sigma - \Delta q_8)~,
\label{s}
\end{equation}
we see that the result for $~\Delta\bar{s}~$ should not change as
$\lambda$ is varied. This provides a serious test for the
stability of the analysis.

Further the dependence of the valence densities upon $\lambda$ is
given by
\begin{eqnarray}
\nonumber
\Delta u_v&=&{1\over 2}[\Delta q_3 + \Delta q_8 -
4(\lambda-1)\Delta\bar{s}]~,\\
\Delta d_v&=&{1\over 2}[-\Delta q_3 + \Delta q_8 - 
4(\lambda-1)\Delta\bar{s}]~,
\label{uvdv}
\end{eqnarray}
so that they are sensitive to the assumption about the sea. On
the other hand, if the analysis is correct, neither 
$~\Delta q_{3,8}~$ nor $~\Delta \Sigma(\Delta\bar{s})~$ nor 
$~\Delta G~$ should change as $\lambda$ is varied.

It should be noticed that Eqs. (\ref {SU3br}) and (\ref {uvdv}) 
are only valid at $Q^2=Q^2_0$. The equality (\ref {SU3br}) and 
therefore Eq. (\ref {uvdv}) will be (marginally) broken because of 
the different NLO evolution of the different sea quarks for 
$Q^2>Q^2_0$.   

It has sometimes been claimed in the literature \cite {MStr}
that as a result of the fits the sea always turns out to be
flavour symmetric. 
Clearly, from the above, one can learn absolutely nothing from the
data about $~\Delta\bar{u}~$ and $~\Delta\bar{d}~$ and any $~\chi^2$
dependence on $\lambda$ must be an artifact of the fitting procedure.\\

{\it 2.3. Simplifying assumptions}\\

In order to limit the number of parameters being fitted, it is
always necessary to take simple functional forms for the parton 
densities, and sometimes additional simplifying assumptions are
made. We have already discussed in the previous section the
assumptions regarding the sea quarks (see Eqs. (\ref{SU3}) and 
(\ref{SU3br})) widely used in the literature. Another somewhat
arbitrary one (for its more recent use see Altarelli et al. 
\cite {Alta}), is to assume that  
\begin{equation}
\Delta q_3(x, Q^2) = C\Delta q_8(x, Q^2)~,
\label{q3q8}
\end{equation}
where $C$ is a constant.

This is, of course, perfectly compatible with the evolution
equations, decreases substantionally the number of parameters
being fitted, but has no physical justification at all.
This assumption leads to a better determination of the rest of
the independent parton densities, but it is not clear whether
the values and errors of the parameters are not thereby
distorted. It is
important to check to what extent (\ref{q3q8}) is compatible with
the results of other theoretical analyses of the data in which 
this approximation is not used.\\ 

{\it 2.4. Scheme dependence}\\

It is well known that at NLO and beyond, the parton densities
become dependent upon the renormalization (or factorization)
scheme. In the unpolarized case the most commonly used are the
$\overline{MS}$, MS and DIS schemes and parton densities in 
different schemes differ from each other by terms of order
$\alpha_s(Q^2)~$, which goes to zero as $Q^2$ increases.

There are two significant differences in the polarized case:

{\it ~i)} The singlet densities $\Delta \Sigma(x, Q^2)$, in two 
different schemes, will differ by terms of order
\begin{equation}
\alpha_s(Q^2)\Delta G(x,Q^2)~,
\label{aldelG}
\end{equation}
which appear to be of order $\alpha_s$. But it is known 
\cite{Efr, CaCoMu} that, as a consequence of the axial anomaly, 
the first moment 
\begin{equation}
\int _{0}^{1} dx\Delta G(x,Q^2) \propto [\alpha_s(Q^2)]^{-1}~,
\label{delG1}
\end{equation}
grows in such a way with $Q^2$ as to compensate for the factor
$\alpha_s(Q^2)$ in (\ref{aldelG}). Thus the difference between 
$\Delta \Sigma$ in different schemes is only apparently of order 
$\alpha_s(Q^2)$, and could be quite large.

{\it ii)} Because of ambiguities in handling the renormalization
of operators involving $\gamma_5$ in $n$ dimensions, the specification
$\overline{MS}$ does {\it not} define a unique scheme. Really
there is a family of $\overline{MS}$ schemes which, strictly,
should carry sub-label indicating how $\gamma_5$ is handled. What
is now conventionally called $\overline{MS}$ is in fact the
scheme due to Vogelsang and Mertig and van Neervenen \cite
{nlocor}, in which the first moment of the non-singlet densities
is conserved, i.e. is independent of $Q^2$, corresponding to the
conservation of the non-singlet axial-vector Cabibbo currents.

Although mathematically correct it is a peculiarity of this
factorization scheme that certain soft contributions are included
in the Wilson coefficient functions, rather than being absorbed
completely into the parton densities. As a consequence, the first
moment of $\Delta \Sigma$ is not conserved so that it is difficult
to know how to compare the DIS results on $\Delta \Sigma$ with
the results from constituent quark models at low $Q^2$.

To avoid these idiosyncrasies Ball, Forte and Ridolfi
\cite{anomdg} introduced what they called the AB scheme, which
involves a minimal modifications of the $\overline{MS}$ scheme,
and for which
\begin{eqnarray}
\nonumber
\Delta \Sigma(x,Q^2)_{AB}&=&\Delta \Sigma(x,Q^2)_{\overline{MS}}+
N_f{\alpha_s(Q^2)\over 2\pi}
\int _{x}^{1} {dy\over y}\Delta G(y,Q^2)_{\overline{MS}}~,\\
\Delta G(x,Q^2)_{AB}&=&\Delta G(x,Q^2)_{\overline{MS}}
\label{ABMS}
\end{eqnarray}
or, in the Mellin $n$-moment space, 
\begin{eqnarray}
\nonumber
\Delta \Sigma(n,Q^2)_{AB}&=&\Delta \Sigma(n,Q^2)_{\overline{MS}}+
N_f{\alpha_s(Q^2)\over 2\pi n}
\Delta G(n,Q^2)_{\overline{MS}}~,\\
\Delta G(n,Q^2)_{AB}&=&\Delta G(n,Q^2)_{\overline{MS}}~.
\label{ABMSmom}
\end{eqnarray}

That $\Delta \Sigma(n=1)_{AB}$ is independent of $Q^2$ to all orders
follows from the Adler-Bardeen theorem \cite{AdlBard}.

The singlet part of the first moment of the structure function $g_1$
\begin{equation}
\Gamma^{(s)}_1(Q^2)\equiv \int _{0}^{1} dxg_1^{(s)}(x,Q^2)
\label{Gam1s}
\end{equation}
then depends on $\Delta \Sigma$ and $\Delta G$ only in the
combination 
\begin{equation}
a_0(Q^2)=\Delta \Sigma(1,Q^2)_{\overline{MS}}=
\Delta \Sigma(1)_{AB}-N_f{\alpha_s(Q^2)\over 2\pi}
\Delta G(1,Q^2)
\label{a0}
\end{equation}
and the unexpectedly small value for the axial charge $a_0$ found 
by the EMC
\cite{EMC}, which triggered the "spin crisis in the parton model"
\cite{ELA}, can be nicely explained as due to a cancellation
between a reasonably sized $\Delta \Sigma(1)$ and the gluon
contribution. Of importance for such an explanation are
both the positive sign and the large value (of order 
${\cal O}(1)$) for the first moment of the polarized gluon density 
$\Delta G(1,Q^2)$ at {\it small} $Q^2\sim 1-10~GeV^2$. Note that
what follows from QCD is that $\vert {\Delta G(1,Q^2)}\vert$ 
grows with $Q^2$ (see Eq. (\ref {delG1})) but its value at 
{\it small} $Q^2$ is unknown in the theory at present and has to 
be determined from experiment.\\ 
  
Although the AB scheme corrects the most glaring weakness of the 
$\overline{MS}$ scheme, it does not consistently put all hard
effects into the coefficient functions. As pointed out in 
\cite{Zijlstra}
one can define a family of schemes labelled by a parameter $a$:
\begin{equation}
\pmatrix {\Delta \Sigma \cr \Delta G}_{a}=     			 
\pmatrix {\Delta \Sigma \cr \Delta G}_{\overline{MS}} 
+{\alpha_{s}\over 2\pi}\pmatrix{0  &z(a)_{qG} \cr
                                0  &0}\otimes
\pmatrix {\Delta \Sigma \cr \Delta G}_{\overline{MS}}
\label{afamily}
\end{equation}
where
\begin{equation}
z_{qG}(x;a) = N_f[(2x-1)(a-1)+2(1-x)]~,
\label{zqG}
\end{equation}
in all of which (\ref{a0}) holds, but which differ in their
expression for the higher moments. (The AB scheme corresponds to
taking $a=2$).

Amongst these we believe there are compelling reasons to choose
what we shall call the JET scheme ($a=1$), i.e.
\begin{equation}
z^{JET}_{qG} = 2N_f(1-x)~.
\label{zjet}
\end{equation}

\def\thefootnote{\dagger}
This is the scheme originally suggested by Carlitz, Collins and
Mueller \cite{CaCoMu} and also advocated by Anselmino, Efremov
and Leader \cite{AEL}.\footnote{There is misprint in Eq. (8.2.6)
of \cite{AEL}. The term $ln({1-x/x^{\prime}\over x/x^{\prime}})$ 
should be
$[ln({1-x/x^{\prime}\over x/x^{\prime}})-1]$.} In it all hard 
effects are absorbed into the coefficient functions. In this scheme 
the gluon coefficient function is exactly the one that would appear 
in the cross section for 
\begin{equation}
pp\rightarrow jet({\bf k}_{T}) + jet(-{\bf k}_{T})+X~,
\label{2jetprod}
\end{equation}
i.e., the production of two jets with large transverse momentum
${\bf k}_{T}$ and $-{\bf k}_{T}$, respectively.

More recently M\"{u}ller and Teryaev \cite{MulTer} have advanced
rigorous and compelling arguments, based upon a generalization of
the axial anomaly to bilocal operators, that removal of all
anomaly effects from the quark densities leads to the JET scheme.
Also a different argument by Cheng \cite{Cheng} leads to the same
conclusion. (Cheng calls the JET scheme a chirally invariant (CI)
scheme.)

The transformation from the $\overline{MS}$ scheme of Mertig, van
Neerven and Vogeslang to the JET scheme is given in moment space
by  
\begin{eqnarray}
\nonumber
\Delta \Sigma(n,Q^2)_{JET}&=&\Delta \Sigma(n,Q^2)_{\overline{MS}}+
2N_f{\alpha_s(Q^2)\over 2\pi n(n+1)}
\Delta G(n,Q^2)_{\overline{MS}}~,\\
\Delta G(n,Q^2)_{JET}&=&\Delta G(n,Q^2)_{\overline{MS}}~.
\label{JETMS}
\end{eqnarray}

Of course, (\ref{ABMSmom}) and (\ref{JETMS}) become the same for $n=1$.\\

In this paper we carry out the fitting
procedure in the $\overline{MS}$ scheme and the results can then
be transformed to the other schemes via (\ref{ABMS}) and (\ref{JETMS}).  
However it will be important to carry out the fitting {\it in the
other schemes} as a check on the stability of the whole analysis
\cite{preparation}.\\

{\bf 3. Method of analysis and input parton distributions }
\vskip 4mm
The spin dependent structure function of interest, $~g^N_1(x,Q^2)~$,
is a linear combination of the asymmetries $A^N_{\parallel}$ and 
$A^N_{\bot}$ (or the related virtual photon-nucleon asymmetries 
$A^N_{1,2}$) measured 
with the target polarized longitudinally or perpendicular to the
lepton beam, respectively. Neglecting as usual the
subdominant contributions (see for example \cite{SMCp94}), 
$~A_1^{N}(x,Q^2)~$ can be expressed {\it via} the polarized 
structure function $~g_1^{N}(x,Q^2)~$ as
 
\begin{equation}
A_1^{N}(x,Q^2)\cong (1+\gamma^2){g_1^{N}(x,Q^2)\over F_1^{N}
(x,Q^2)}={g_1^{N}(x,Q^2)\over F_2^{N}(x,Q^2)}[2x(1+R^{N}(x,Q^2)]~, 
\label{assym}
\end{equation} 
where
\begin{equation}
R^N + 1 = (1+\gamma^2)F_2^N/2xF^N_1
\label{ratio}
\end{equation} 
and $~F_1^N~$ and $~F_2^N~$ are the unpolarized structure functions.
In (\ref{assym}) the kinematic factor $\gamma^2$ is given by
\begin{equation}
\gamma^2={4M^2_{N}x^2\over Q^2}~.
\label{g2}
\end{equation}
It should be noted that in the SLAC kinematic region $\gamma$
cannot be neglected.

In some cases the theoretical analyses of the data are presented in
terms of $~g_1^{N}(x,Q^2)~$ as extracted from the measured values of
$~A_1^{N}(x,Q^2)~$ according to (\ref{assym}), using various
parametrizations of the experimental data for $F_2$ and $R$.  

As in our previous analysis we follow the approach first used in 
\cite{Vog}, in which the next-to-leading (NLO) QCD predictions
for the spin-asymmetry $~A_1^{N}(x,Q^2)~$ are confronted with the
data on $~A_1^{N}(x,Q^2)~$, rather than with the $~g_1^{N}(x,Q^2)~$ 
derived by the procedure mentioned above.
The choice of $A^N_1$ should minimize the higher twist
contributions which are expected to partly cancel in the ratio 
(\ref{assym}), allowing use of data at lower $Q^2$.
Bearing in mind that in polarized DIS most of the small $x$ data points 
are at low $~Q^2~$, a lower than usual cut is needed ($~Q^2>1~GeV^2~$) 
in order to have enough data for the theoretical analysis. We believe 
that in this approach such a low $~Q^2$-cut is more justified.\\

In NLO approximation 
\begin{equation}
A_1^{N}(x,Q^2)_{NLO}\cong (1+\gamma^2){g_1^{N}(x,Q^2)_{NLO}\over 
F_1^{N}(x,Q^2)_{NLO}}~.
\label{NLOas}
\end{equation} 

In (\ref{NLOas}) $~N=p,~n~$ and $~d=(p+n)/2~$.

To calculate $~A_1^{N}(x,Q^2)_{NLO}~$ in NLO QCD and then fit  
the data we follow the same procedure described in detail
in our paper \cite{LSiSt}. Here we will recall only the main 
points.

The $Q^2$ evolution, in NLO QCD approximation, is carried out for 
the n-space moments
of the polarized quark and gluon densities. Then using
the known NLO expressions for the moments of the Wilson coefficients 
$\delta C_q(n,\alpha_s)~$ and $~\delta C_G(n,\alpha_s)~$ (see e.g.
\cite{Vog}) one can 
calculate the moments of the structure function  
\begin{equation}
M^N(n,Q^2)={1\over 2}\sum _{q} ^{N_f}e_{q}^2\left
[\delta C_q(n)(\Delta q(n,Q^2)+\Delta\bar{q}(n,Q^2)) +
{1\over N_{f}}\delta C_G(n)\Delta G(n,Q^2) \right].
\label{g1Nmom}
\end{equation}

As already mentioned above, all calculations are performed in the
$\overline{MS}$ scheme. To account for heavy quark contributions 
we use the so-called fixed-flavour scheme \cite{GRV, Vog} and set
the number of active flavours in (\ref{g1Nmom}) $N_f=3$. In
contrast to our previous analysis \cite{LSiSt}, we now use for
the values of the QCD parameter $~\Lambda_{\overline{MS}}~$: 
$~\Lambda_{\overline{MS}}(n_f=3)=353~MeV$ and 
$~\Lambda_{\overline{MS}}(n_f=4)=300~MeV~$, which correspond to
$~\alpha_s(M^2_{z}) = 0.1175~$, as obtained by the MRST analysis
\cite{MRST} of the world unpolarized data, in
excellent agreement with the world average $~\alpha_s(M^2_{z}) =
0.118 \pm 0.005~$ \cite{alsaver}.

Finally, to reconstruct the spin structure functions $~g^{N}_1(x,Q^2)~$
in Bjorken x-space from their moments (\ref{g1Nmom}) with the
required accuracy, we use the Jacobi reconstruction method
\cite{Parisi, Kriv}. Note that in this method the 
structure functions are given analytically.

The same procedure has been used to calculate the unpolarized
structure functions $~F_1^{N}(x,Q^2)_{NLO}~$ from their moments.\\

We choose the input polarized densities at $Q^2_0=1~GeV^2$ in the
form:
\begin{eqnarray}
\nonumber
x\Delta u_v(x,Q^2_0)&=&\eta_u A_ux^{a_u}xu_v(x,Q^2_0)~,\\
\nonumber
x\Delta d_v(x,Q^2_0)&=&\eta_d A_dx^{a_d}xd_v(x,Q^2_0)~,\\
\nonumber
x\Delta Sea(x,Q^2_0)&=&\eta_S A_Sx^{a_S}xSea(x,Q^2_0)~,\\
x\Delta G(x,Q^2_0)&=&\eta_g A_gx^{a_g}xG(x,Q^2_0)
\label{classic}
\end{eqnarray}
where on R.H.S. of (\ref{classic}) we have used the MRST 
unpolarized densities \cite{MRST}. 

Guiding arguments for such an ansatz are simplicity (not too
many free parameters) and the expectation that polarized and
unpolarized densities have similar behaviour at large $x$. 
In (\ref{classic}) 
the parameters $\a_f$ account for the difference of the low-$x$
behaviour between the polarized and unpolarized parton densities. 
The normalization factors 
$~A_f~$ are determined in such a way as to ensure that the first 
moments of the polarized densities are given by $~\eta_{f}$. 

In the previous section we explained why we chose to
deal with valence and sea quarks instead of their singlet and
non-singlet combinations. For the polarized light and strange sea
quark densities at $Q^2_0=1~GeV^2$ we adopt the assumption
(\ref{SU3br}). Then
\begin{equation}
\Delta\bar {s}\equiv \Delta\bar{q}={\Delta Sea\over 2(2\lambda+1)}~,
~~~~~~~\eta_{\bar {s}}={\eta_S\over 2(2\lambda+1)}~,
\label{dels}
\end{equation}
where $\eta_{\bar {s}}$ is the first moment of the strange sea
parton density $\Delta\bar {s}$.

We would like to emphasize once more that in contrast to the valence
quark densities, $\Delta\bar {s}$ should not depend on the flavour
sea decomposition, i.e. on $\lambda$ in our case
(see Eq. (\ref{s})), and as will be seen below, our numerical results 
confirm this.
 
The first moments of the valence quark  densities
$~\eta_u~$ and $~\eta_d~$ are fixed by the octet hyperon $\beta$ 
decay constants \cite{PDG}
\begin{equation}
g_{A}=F+D=1.2573~\pm~0.0028,~~~a_8=3F-D=0.579~\pm~0.025~.
\label{GA3FD}
\end{equation}
to be
\begin{equation}
\eta_u=0.918-2(\lambda-1)\Delta\bar {s}(1,Q^2_0),~~~~
\eta_d=-0.339-2(\lambda-1)\Delta\bar {s}(1,Q^2_0).
\label{etaud}
\end{equation}
In the case of SU(3) flavour symmetry of the sea $(\lambda=1)$
we take
\begin{equation}
\eta_u=0.918~,~~~~~~~~\eta_d=-0.339~.
\label{etaudSU3}
\end{equation}

The rest of the parameters in (\ref{classic})
\begin{equation}
\{a_u,~a_d,~\eta_S,~a_S~,\eta_g~,a_g\}~,
\label{claspar}
\end{equation}
have to be determined from the best fit to the $~A_1^N(x,Q^2)~$ data.

In some papers \cite{Alta, Bour} $g_A$, and in others (see e.g. 
\cite{E154th}) both $g_A$ and $a_8$ have been taken to be free
parameters determined by the best fit to the inclusive polarized 
DIS data. We do not favour such an approach because the values
of these quantities, especially $g_A$, are determined from much
more precise experiments and using them improves the accuracy
with which we can determine the polarized densities.\\

{\bf Results of Analysis}
\vskip 4mm

In this section we present the results of our fits to the
present experimental data on
$~A_1^N(x,Q^2)~$: EMC proton data \cite{EMC}, SLAC E142
neutron data \cite{SLACnQ2}, SLAC E154 neutron data 
{\cite{E154}, SMC combined proton data \cite{newSMCpQ2}, 
the SMC deuteron data \cite{SMCd97Q2}
which are combined data from the 1992, 1994 \cite{SMCd93_95}
and 1995 runs, HERMES neutron data \cite{HERMES} and the final 
SLAC E143 results \cite{finSLACpd}
on $~g_1^p/F_1^p~$ and $~g_1^d/F_1^d~$.
The data used (354 experimental points) cover the following 
kinematic region:  
\begin{equation}
0.004< x < 0.75,~~~~~~1< Q^2< 72~GeV^2~.
\label{kinreg}
\end{equation}

As mentioned in the Introduction, in contrast to the other analyses,
we fit all possible $(x,Q^2)$ data rather than ones
averaged over $Q^2$ within each $x$-bin. We denote this combined fit 
to the $(x,Q^2)$  data on $A_1^N$ presented by E142, E143 and SMC
collaborations and the averaged $A_1^N$ data given by EMC, E154
and HERMES as Fit A. Since for most of these
data (E142, SMC) the systematic errors are not published, in 
Fit A only statistical errors are taken into account. 
The results of the fit to the averaged $~A^{N}_1~$ data alone (118
experimental data points) given by all the collaborations
mentioned above are also presented (Fit B). In this case the
total (statistical and systematic) errors are included in the
analysis. "Higher twist" corrections are not
included in the present study. As already discussed
above, in the approach used their effect is expected to be 
negligible.\\  

The numerical results of the fits ($\lambda =1$) are listed in
Table 1. Note that $\eta_{\bar{s}}=\eta_S/6~$ in the SU(3)
symmetry case. The dependence of the results on the flavour
decomposition of the sea will be discussed below in detail. 

It follows from our analysis that the value of $a_g$ can not be well
determined, i.e. the existing data do not constrain the behaviour
of the polarized gluon density at small $x$. For that reason the 
fits to the data were performed at different fixed values of $a_g$
in the range:$0\leq a_g\leq 1$. In Fit A the change of 
$~\chi^2/DOF~$
value from $~\chi^2/DOF(a_g=0)=0.917~$ to $~\chi^2/DOF(a_g=1)
=0.910~$ is negligible. The same conclusion is true for the fits
to the average $A_1$ data. In Table 1 we present the results of
the fits corresponding to $a_g=0.6$.\\

In Fig. 1 the SLAC/E143 and SMC data on $A^p_1$ and $A^d_1$ vs
$Q^2$ for different $x$-bins are compared to our best Fit A. The
NLO results for the averaged asymmetries $~A^N_1~$ (Fit B) are
shown in Fig. 2. It is seen from the values of $\chi^2/DOF$ and
Figs. 1 and 2 that the NLO QCD predictions are in a very good
agreement with the presently available data on $A^N_1$, as well
as with the corresponding $g_1^N(x, Q^2)$ data (see Figs. 6a and b). 
We would like to draw special attention to the excellent fit to the
E154 neutron data (see Fig. 2c), the most accurate polarized DIS 
data at present ($\chi^2=1.6$ for 11 experimental data points). 
\vskip 0.6 cm
\begin{center}
\begin{tabular}{cl}
&{\bf Table 1.} Results of the NLO QCD fits to the world 
$~A_1^N~$ data ($Q^2_0=1~GeV^2$).\\ 
&For Fit A errors are statistical, for Fit B, total.
$a_g=0.6$ (fixed).
\end{tabular}
\vskip 0.6 cm
\begin{tabular}{|c|c|c|c|c|c|c|} \hline
 ~~Parameters~~  &~~~~~~~~~~~~~~~Fit A~~~~~~~~~~~~~~~
 &~~~~~~~~~~~~~~~Fit B~~~~~~~~~~~~~~~\\ \hline
 $DOF$           &  354~-~5      &     118~-~5 \\
 $\chi^2$        &  318.4        &     86.1    \\
 $\chi^2/DOF$    &  0.912        &     0.762   \\  \hline
 $a_u$           &~~0.250~~$\pm$~~0.023 &~~0.255~~$\pm$~~0.028\\
 $a_d$           &~~0.231~~$\pm$~~0.088 &~~0.148~~$\pm$~~0.113\\
 $a_S$           &~~0.576~~$\pm$~~0.152 &~~0.817~~$\pm$~~0.223\\
 $\eta_{\bar{s}}$&-~0.054~~$\pm$~~0.012 &-~0.049~~$\pm$~~0.005\\
 $\eta_g$        &~~~0.34~~$\pm$~~0.24  &~~~0.82~~$\pm$~~0.32~~\\ \hline
 $a_0=\Delta \Sigma(1)_{\overline{MS}}$&~~0.253~~$\pm$~~0.079
 &~~0.287~~$\pm$~~0.041\\ 
 $\Delta \Sigma(1)_{AB}$&~~0.332~~$\pm$~~0.096&~~0.476~~$\pm$~~0.084\\  
\hline
\end{tabular}
\end{center}
\vskip 0.6 cm

One can see from Fig. 1 that the accuracy and the presently measured 
kinematic region of the data do not allow a definite
conclusion about the scaling violations in $A_1^N(x, Q^2)$. It is
obvious that more precise data and an extension of the measured
range to smaller $x$ and larger $Q^2$ are needed to answer the
question about the $Q^2$ dependence of virtual photon spin asymmetry 
$A^N_1$.\\

The extracted valence, strange and gluon polarized distributions
at $Q^2=1~GeV^2$ are shown in Fig. 3a: Fit A (solid curves) and
Fit B (dashed curves). Except for the gluon these two sets of
densities coincide almost exactly in the measured $x$-region. 
The polarized gluon densities
extracted in Fit A and Fit B are a good illustration of how large 
the uncertainty is in determining the gluon density from the present 
data. Although the values of their first moments $\eta_g$ are in
agreement within two standard deviations, their central values
differ by a factor of more than two.
(Note that in Fit A and B the same form of the initial 
parton densities has been used.) 

In Fig. 3b  and Fig. 4 the nonsinglet combinations $x\Delta q_3~,
x\Delta q_8$ and their ratio $\Delta q_3/\Delta q_8$ are shown,
respectively. The ratio $\Delta q_3/\Delta q_8$ is shown at 
$Q^2=1~GeV^2$, but it does not change with $Q^2$ because the same
evolution holds for both $x\Delta q_3$ and $x\Delta q_8$. 
We find a significant deviation from the approximation 
(\ref{q3q8}) used in \cite{Alta} by Altarelli et al. Note that 
the $Q^2$ evolution of the nonsinglets is the same in all the 
schemes discussed in Section 2.4.\\

Let us now examine how the assumption (\ref{SU3br}) about the flavour 
decomposition of the sea influences our results. As pointed out
in Section 2.2 the strange quark density, and in particular, its 
first moment $\Delta\bar{s}(1,Q^2_0)$ should not change as 
$\lambda$ is varied. The results of the fits to the averaged $A_1^N$
data (Fit B) using different values of $\lambda$ are presented in 
Table 2.
\vskip 0.6 cm
\begin{center}
\begin{tabular}{cl}
&{\bf Table 2.} The results for the first
moments of the polarized distributions at\\ 
&$Q^2=1~GeV^2$ using the assumption (\ref{SU3br}) about 
the flavour decomposition of\\ 
&the sea.
\end{tabular}
\vskip 0.6 cm
\begin{tabular}{|c|c|c|c|c|c|c|} \hline
~~$\lambda$~~&~~$\chi^2$~~&$-~\eta_{\bar{s}}$ &
 $\eta_g$ & $\eta_u$ &$-~\eta_d$\\ \hline
  0.5 & 85.97 &0.049~$\pm$~0.004 & 0.78~$\pm$~0.31 & 
  0.869~$\pm$~0.013 & 0.388~$\pm$~0.013\\
  1.0 & 86.11 &0.049~$\pm$~0.005 & 0.82~$\pm$~0.32 &
  0.918~$\pm$~0.013 & 0.339~$\pm$~0.013\\
  2.0 & 86.08 & 0.050~$\pm$~0.007 & 0.86~$\pm$~0.34 & 
  1.018~$\pm$~0.019 & 0.239~$\pm$~0.019\\
  3.0 & 86.12 & 0.048~$\pm$~0.004 & 0.93~$\pm$~0.35 & 
  1.110~$\pm$~0.020 & 0.147~$\pm$~0.020 \\ \hline
\end{tabular}
\end{center}
\vskip 0.6 cm

It is clear from the table that $\chi^2$ and the central values of
$\eta_{\bar{s}}\equiv \Delta\bar{s}(1,Q^2_0)$ and $\eta_g\equiv
\Delta G(1,Q^2_0)$ practically do not change
as $\lambda$ varies. We regard this fact as a very good test
of the stability of our analysis. We thus conclude, somewhat 
in disagreement with Ref. \cite{Alta} that the separation into
valence and sea contributions need not introduce biases into the fit
provided sufficient  care is taken. Of course, we have also 
demonstrated very clearly that the $\Delta u_v$ and $\Delta d_v$ 
valence quark densities are sensitive to the assumptions about 
the sea. This dependence is shown in Fig. 5.     

What follows from our analysis is that one can use for input
distributions a parametrization in terms of valence and sea
quarks. With a correct fitting procedure the strange sea (or
singlet, see (\ref{s})) and gluon distributions defined from the
inclusive data do not depend on the assumption about the sea and
therefore, one can use them to test the remarkable relation 
(\ref{a0}). On the contrary, as emphasized in Section 2.2, 
electromagnetic DIS does not fix the 
valence quark densities, which are sensitive to the assumption 
about the sea. This implies that is somewhat meaningless to claim 
\cite{deFlorian} that
"the present semi-inclusive data alone fail to define a $\Delta d_v$
distribution consistent with those extracted from inclusive data".\\

In Table 1 we also present our results for the first moments of the 
polarized gluon and singlet quark densities in the $\overline{MS}$
(determined from the fit) and AB (calculated by Eq. (\ref{ABMSmom}))
schemes. The value of $\eta_g$ (Fit B)
\begin{equation}
\eta_g\equiv \Delta G(1,1~GeV^2)_{\overline{MS}}=
\Delta G(1,1~GeV^2)_{AB}=0.82\pm 0.32
\label{etag}
\end{equation}
is in a good agreement with the one obtained in Fit A of 
Ref. \cite{Alta}:
\begin{equation}
\Delta G(1,1~GeV^2)_{AB}=0.95\pm 0.18
\label{etagAB}
\end{equation}
using almost the same data.

Our result for $\Delta \Sigma(1)_{AB}$
\begin{equation}
\Delta \Sigma(1)_{AB}=0.476\pm 0.084
\label{sig1ABour}
\end{equation}
is in agreement within two standard deviations with its
constituent value 0.6 \cite{Jaffe} and, within errors, coincides
with the value determined in \cite{Alta}
\begin{equation}
\Delta \Sigma(1)_{AB}=0.405\pm 0.032~.
\label{sigABAlta}
\end{equation}

Finally, let us turn to the first moments $\Gamma^{N}_1(Q^2)$ 
of the spin structure function $g_1^{N}(x, Q^2)$. Using our
results of the fits for the input polarized parton densities
these quantities have been calculated for different values of 
$Q^2$ using Eq. (\ref{g1Nmom}) for $n=1$. The results are 
presented in Table 3.
\vskip 0.6 cm
\begin{center}
\begin{tabular}{cl}
&{\bf Table 3.} Determination of the first moments
$\Gamma^{N}_1(Q^2)$ of $g_1^{N}(x, Q^2)$\\
&using the results of our NLO QCD fits to the world 
$~A_1^N~$ data. 
\end{tabular}
\vskip 0.6 cm
\begin{tabular}{|c|c|c|c|c|c|c|} \hline
               &~~~~~~~~~~Fit A~~~~~~~~~~
 &~~~~~~~~~~Fit B~~~~~~~~~~\\ \hline
 $\Gamma^{p}_1(5~GeV^2)$&~~0.134~~&~~0.138~~\\
 $\Gamma^{n}_1(5~GeV^2)$&-~0.056~~&-~0.053~~\\
 $\Gamma^{d}_1(5~GeV^2)$&~~0.036~~&~~0.039~~\\
 $\Gamma^{p}_1(10~GeV^2)$&~~0.136~~&~~0.139~~\\
 $\Gamma^{n}_1(10~GeV^2)$&-~0.057~~&-~0.054~~\\
 $\Gamma^{d}_1(10~GeV^2)$&~~0.036~~&~~0.039~~\\
 \hline
\end{tabular}
\end{center}
\vskip 0.6 cm

The values of the first moments of the polarized parton densities
and therefore, the corresponding values of the first moments of
$g_1^N$ are very sensitive to the assumed small $x$-behavior of
the input parton densities (see e.g. \cite{E154, newSMCpQ2}).
Parton densities which give the same results for the physical
quantities in the {\it measured} $x$ region can lead to rather
different behaviour at very small $x$. The dependence of the 
the moments of the physical quantities in polarized DIS on 
different assumptions
about the small $x$ behaviour of the input parton densities has
been studied in detail in \cite{Alta}.

The simple Regge behaviour of the unpolarized and polarized
structure functions as $x\rightarrow 0$, which was for many years
a guiding principle in our choice of the starting ansatz for the
parton densities, differs from that predicted in QCD \cite{QCDsmallx,
Ryskin}. Moreover, the unpolarized DIS experiments at HERA have
shown that the simple Regge extrapolation of the structure
functions at small $x$ is not valid at large $Q^2$. There is
an indication \cite{E154} that this is also true in the polarized
case. The question of the small $x$ behaviour in the polarized
case remains an open one and is a serious challenge to both
experiment and theory.

For illustration we compare our predictions for  
$\Gamma^{p}_1(10~GeV^2)$ and  $\Gamma^{n}_1(5~GeV^2)$
\begin{eqnarray}
\nonumber
\Gamma^{p}_1(10~GeV^2)&=&\{0.136~(Fit~A),~~0.139~(Fit~B)\}~,\\
\Gamma^{n}_1(5~GeV^2)&=&\{-0.057~(Fit~A),-0.054~(Fit~B)\}
\label{Gam1pn}
\end{eqnarray}
with their "experimental" values \cite{newSMCpQ2} 
and \cite{E154th}
\begin{equation}
\Gamma^{p}_1(10~GeV^2)=0.130\pm 0.006(stat)\pm 0.008(syst)
\pm 0.014(evol)~,
\label{Gam1pexp}
\end{equation}
\begin{equation}
\Gamma^{n}_1(5~GeV^2)=-0.058\pm 0.004(stat)\pm 0.007(syst)
\pm 0.007(evol)~,
\label{Gam1nexp}
\end{equation}
the latter chosen because the moments have been determined by
extrapolating the polarized structure functions into the 
unmeasured $x$ region using QCD. It is seen from Eqs. 
(\ref{Gam1pn})-(\ref{Gam1nexp}) that our predictions for 
$\Gamma^{p}_1$ and $\Gamma^{n}_1$ are in a good agreement with
their "experimental" values.\\

{\bf 5. Conclusion}
\vskip 4mm

We have performed a next-to leading order QCD analysis 
($\overline{MS}$ scheme) of the world data on inclusive 
polarized deep inelastic lepton-nucleon scattering.
The QCD predictions have been confronted with the data on the
virtual photon-nucleon asymmetry $~A_1^N(x,Q^2)~$, rather than
with the polarized structure function $~g_1^N(x,Q^2)~$, in order
to minimize the higher twist effects. In this paper, for the
first time, is utilized the full world data set on $A_1^N$ with 
its detailed binning in $(x,Q^2)$. Using the simple parametrization 
(\ref{classic}) (with only 5 free parameters) for the input 
polarized parton densities it was
demonstrated that the polarized DIS data are in an excellent
agreement with the pQCD predictions for $~A_1^N(x,Q^2)~$ and the 
spin-dependent structure function $~g_1^N(x,Q^2)~$. However, the 
accuracy and the presently measured kinematic region of the data 
do not allow a definite conclusion about scaling violations 
in $A_1^N(x, Q^2)$.

We have studied the consequences of different simplifying 
assumptions, usually
made in recent analyses to aid the extraction of the polarized
parton densities from the data. It was shown
that whereas the valence quark densities determined from 
inclusive polarized DIS data, are sensitive to the assumptions
about the flavour decomposition of the sea, the extracted strange
sea and gluon densities in our analysis do {\it not} depend on 
such an assumption
and can therefore be used to test the remarkable relation
(\ref{a0}). We have found a significant deviation from the
approximation (\ref{q3q8}) used in Ref. \cite{Alta} that the
nonsinglet quark distributions $~\Delta q_3(x,Q^2)~$ and 
$~\Delta q_8(x,Q^2)$ have the same shape at fixed $Q^2$. 

Although the quality of the data has significantly improved via
the recent experiments of the SLAC/E154 Collaboration and the 
final SMC results on $A_1^p$, the uncertainty in determining the
polarized gluon density is still very large.\\

Despite the great progress of the past few years it is clear that 
in order to test precisely the spin properties of QCD, more accurate 
inclusive DIS polarized data and an extension of the measured
range to smaller $x$ and larger $~Q^2~$ are needed. We hope the
current DIS experiments at HERA will help in clarifying the
situation. In addition, semi-inclusive and charged current data 
will be very important for a precise determination of the polarized 
parton densities and especially, for an accurate flavour decomposition 
of the polarized quark sea. There is some progress in this direction
\cite{deFlorian}.
Finally, a direct measurement of $~\Delta G(x,Q^2)~$ in
processes such as $J/\psi$ production in lepton-hadron scattering
with a polarized beam will answer the important question about the
magnitude of the first moment of the gluon density $~\Delta G(x,Q^2$).\\
 
{\bf Acknowledgments}
\vskip 3mm

We are grateful to O. V. Teryaev for useful discussions and remarks.\\

This research was partly supported by a UK Royal Society Collaborative
Grant, by the Russian Fund for Fundamental Research Grant No 
96-02-17435a and by the Bulgarian 
Science Foundation under Contract \mbox{Ph 510.}\\


\newpage
\noindent
{{\bf Figure Captions}}
\vskip 3mm
\noindent 
{\bf Fig. 1a-1b.} $A_1^p$ and $A_1^d$ vs $Q^2$ for different 
$x$-bins for the proton E143 \cite{SLACpdQ2} and SMC data
\cite{newSMCpQ2} and for deuteron E143 \cite{SLACpdQ2} and SMC data 
\cite{SMCd97Q2}.
Only statistical errors are shown. The solid curves correspond to
Fit A described in the text. The E143 data on $g_1/F_1$ are
multiplied by the kinematic factor $(1+\gamma^2)~$.\\ 

\noindent 
{\bf Fig. 2a-2c.} Comparison of our NLO results (Fit B) for 
averaged $~A_1^N(x,Q^2)~$
with the experimental data at the measured $x$ and $Q^2$ values. 
Errors bars represent the total error.\\
 
\noindent 
{\bf Fig. 3a-3b.} Next-to-leading order input polarized parton
distributions at $~Q^2=1~GeV^2~$ ($\lambda=1$). Solid and dashed
curves in Fig. 3a correspond to Fit A and Fit B, respectively.\\

\noindent 
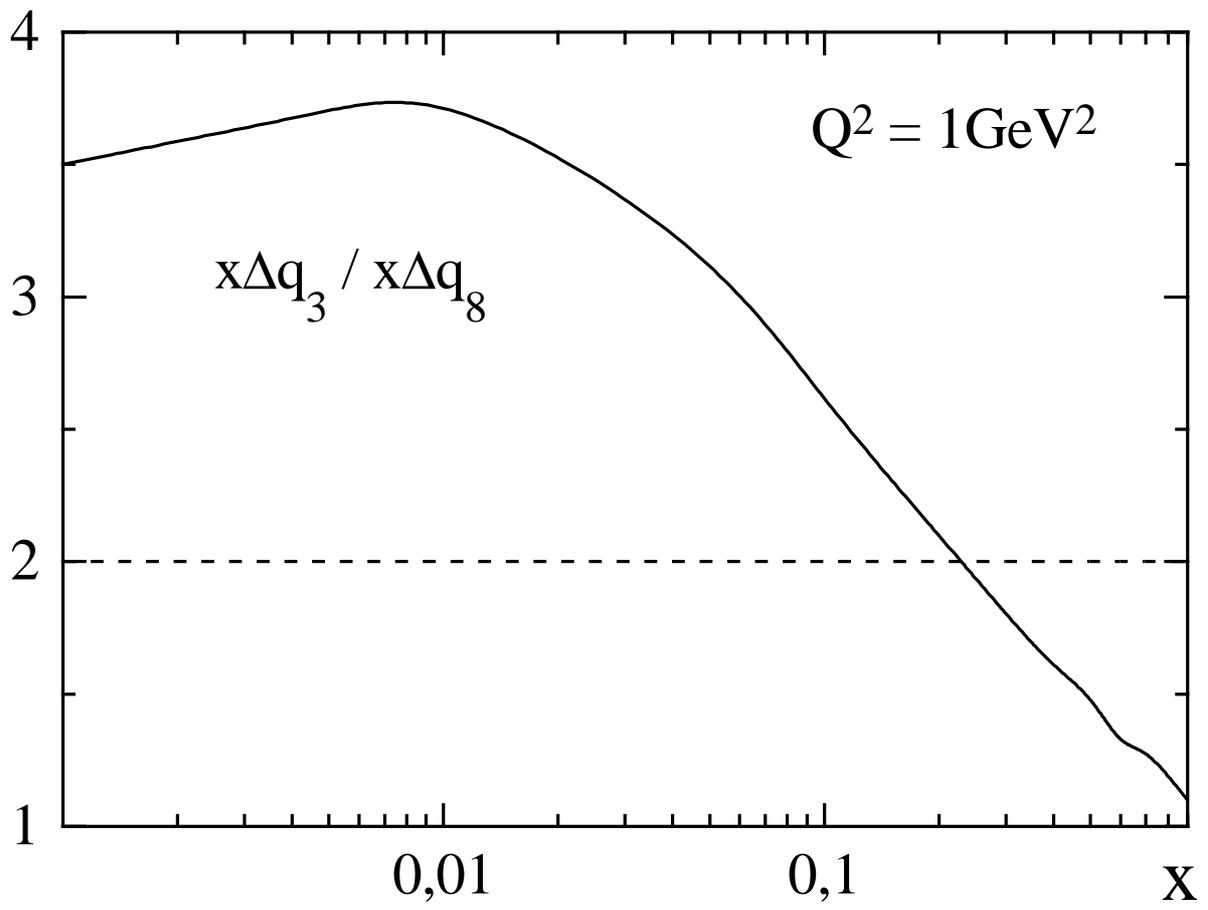
{\bf Fig. 4.} Comparison between our result (solid curve) for the 
ratio $\Delta q_3(x)/\Delta q_8(x)$ and approximation 
(\ref{q3q8}) used in Ref. \cite{Alta} (dashed line).\\
 
\noindent 
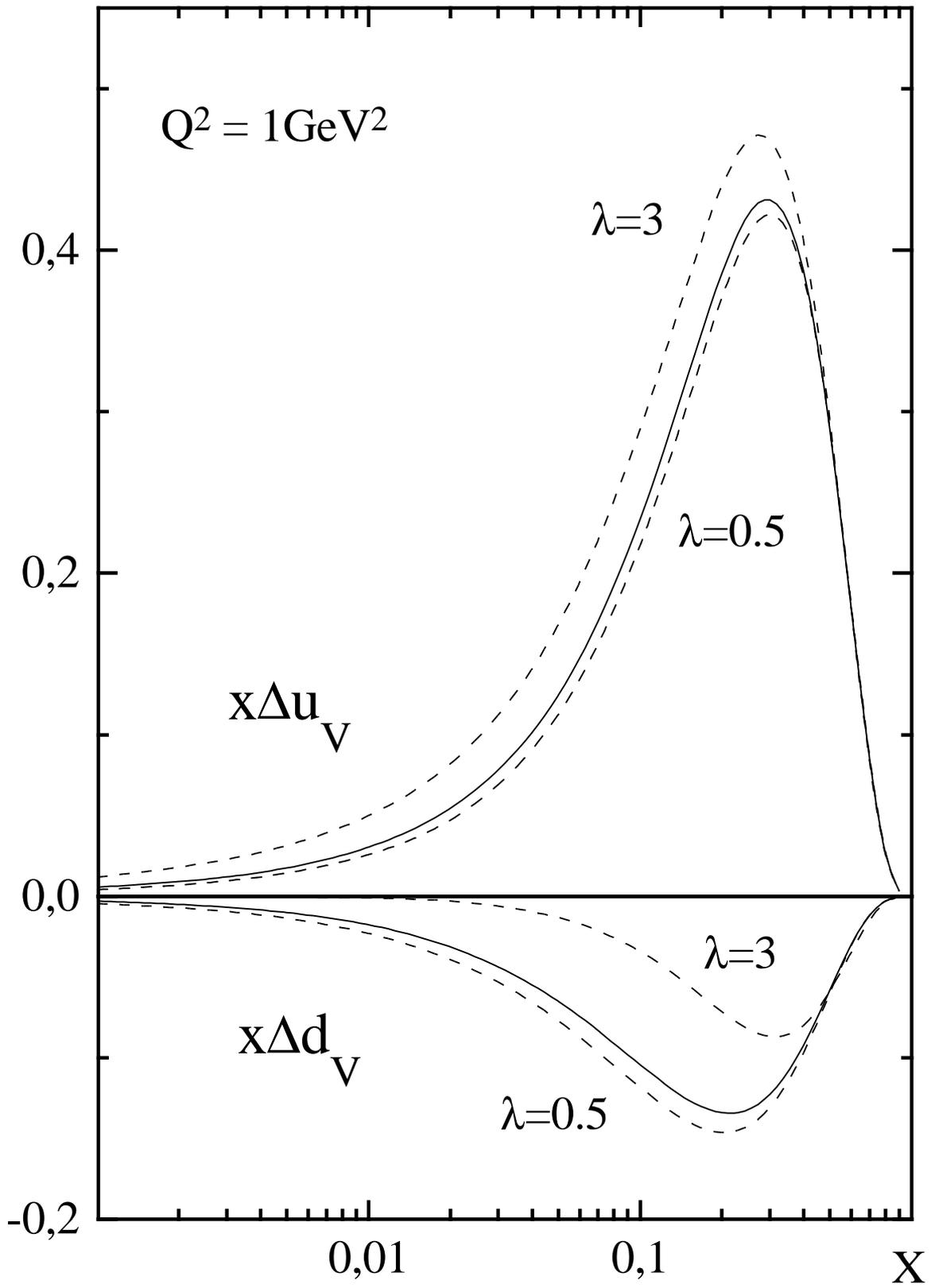
{\bf Fig. 5.} Polarized valence quark densities at $~Q^2=1~GeV^2~$ 
for different values of $\lambda$ (see Eq. (\ref{SU3br})).
The solid curves correspond to an $SU(3)$ flavour symmetric sea 
($\lambda=1$).\\

\noindent 
{\bf Fig. 6a-6b.} Comparison of our NLO results for
$~g_1^N(x,Q^2)~$ (6a) and $~xg_1^N(x,Q^2)~$ (6b) with SMC 
\cite{SMCd97Q2, newSMCpQ2} and SLAC/E154 data \cite{E154}
at the measured values of $Q^2$. Error bars represent the total 
error.


\begin{thebibliography}{99}

\bibitem{EMC}
EMC, J. Ashman et al.,
\newblock {\it Phys. Lett.} {\bf B206}, 364 (1988);\\
{\it Nucl. Phys.} {\bf B328}, 1 (1989).

\bibitem{SLACnQ2}
SLAC E142 Collaboration, P. L. Anthony et al.,
\newblock {\it Phys. Rev.} {\bf D54}, 6620 (1996).

\bibitem{SLACpd}
SLAC E143 Collaboration, K. Abe et al.,
\newblock {\it Phys. Rev. Lett.} {\bf 74}, 346 (1995);\\
{\it Phys. Rev. Lett.} {\bf 75}, 25 (1995).

\bibitem{SLACpdQ2}
SLAC E143 Collaboration, K. Abe et al.,
\newblock {\it Phys. Lett.} {\bf B364}, 61 (1995).

\bibitem{finSLACpd}
SLAC E143 Collaboration, K. Abe et al.,
Preprint SLAC-PUB-7753, Feb 1998, e-Print Archive:hep-ph/9802357.

\bibitem{E154}
SLAC/E154 Collaboration, K. Abe et al., 
\newblock {\it Phys. Rev. Lett.} {\bf 79}, 26 (1997).

\bibitem{SMCd93_95}
SMC, D. Adeva et al.,
\newblock {\it Phys. Lett.} {\bf B302}, 533 (1993);\\
D. Adams et al.,
\newblock {\it Phys. Lett.} {\bf B357}, 248 (1995).

\bibitem{SMCp94}
SMC, D. Adams et al.,
\newblock {\it Phys. Lett.} {\bf B329}, 399 (1994),\\
erratum {\it ibid} {\bf B339} 332 (1994);
{\it Phys. Rev.} {\bf D56}, 5330 (1997).

\bibitem{SMCd97Q2}
SMC, D. Adams et al.,
\newblock {\it Phys. Lett.} {\bf B396}, 338 (1997).

\bibitem{newSMCpQ2}
SMC, D. Adeva et al.,
\newblock {\it Phys. Lett.} {\bf B412}, 414 (1997).

\bibitem{HERMES}
HERMES Collaboration, K. Ackerstaff et al.,
\newblock {\it Phys. Lett.} {\bf B404}, 383 (1997).

\bibitem{g2}
SMC, D. Adams et al.,
\newblock {\it Phys. Lett.} {\bf B336}, 125 (1994);\\
SLAC E143 Collaboration, K. Abe et al.,
\newblock {\it Phys. Rev. Lett.} {\bf 76}, 587 (1996);\\
SLAC/E154 Collaboration, K. Abe et al., 
\newblock {\it Phys. Lett.} {\bf B404}, 377 (1997).

\bibitem{GSt}
T. Gehrmann and W. J. Stirling,
\newblock {\it Phys. Rev.} {\bf D53}, 6100 (1996).

\bibitem{Vog}
M. Gl\"{u}ck, E. Reya, M. Stratmann and W. Vogelsang,
\newblock {\it Phys. Rev.} {\bf D53},\\
 4775 (1996).

\bibitem{anomdg}
R. D. Ball, S. Forte and G. Ridolfi,
\newblock {\it Phys. Lett.} {\bf B378}, 255 (1996).

\bibitem{Alta}
G. Altarelli, R. D. Ball, S. Forte and G. Ridolfi, 
\newblock {\it Nucl. Phys.} {\bf B496}, 337 (1997);\\
{\it Acta Phys. Polon.} {\bf B29}, 1145 (1998),
e-Print Archive:hep-ph/9803237.

\bibitem{E154th}
SLAC/E154 Collaboration, K. Abe et al.,
\newblock {\it Phys. Lett.} {\bf B405}, 180 (1997).

\bibitem{LSiSt}
E. Leader, A. V. Sidorov and D. B. Stamenov,\\
e-Print Archive:hep-ph/9708335 (to be published in IJMPA).

\bibitem{MStr}
M. Stratmann, Preprint DO-TH 97/22, October 1997,\\ 
e-Print Archive:hep-ph/9710379.
  
\bibitem{DeRoeck}
A. De Roeck at al., Preprint DESY 97-249, e-Print Archive:hep-ph/9801300.

\bibitem{Bour}
C. Bourrely, F. Buccella, O. Pisanti, P. Santorelli and J. Soffer,\\
Preprint CPT-97/P 3578, e-Print Archive:hep-ph/9803229.

\bibitem{nlocor}
R. Mertig and W. L. van Neerven,
\newblock {\it Z. Phys.} {\bf C70}, 637 (1996);\\
W. Vogelsang,
\newblock {\it Phys. Rev.} {\bf D54}, 2023 (1996).

\bibitem{MRST}
A. D. Martin, R. G. Roberts, W. J. Stirling and R. S. Torn,\\
Preprint RAL-TR-98-029, e-Print Archive:hep-ph/9803445.

\bibitem{alsaver}
See for example:W. J. Stirling, Preprint DTP-97-80, Jun 1997,\\
e-Print Archive:hep-ph/9709429.

\bibitem{semiincl}
EMC, J. Ashman et al.,
\newblock {\it Nucl. Phys.} {\bf B328}, 1 (1989);\\
SMC Collaboration, B. Adeva et al.,
\newblock {\it Phys. Lett.} {\bf B369}, 93 (1996);\\
{\bf B420}, 180 (1998).

\bibitem{deFlorian}
D. de Florian, O. A. Sampayo and R. Sassot,
\newblock {\it Phys. Rev.} {\bf D57}, 5803 (1998).

\bibitem{Efr}
A. V. Efremov and O. V. Teryaev, Dubna report E2-88-287, 1988
(published in the Proceedings of the Int. Hadron Symposium,
Bechyne, Czechoslovakia, 1988, eds. J. Fischer et al. (Czech
Academy of Science, Prague, 1989), p. 302);\\
G. Altarelli and G. G. Ross, 
\newblock {\it Phys. Lett.} {\bf B212}, 381 (1988).

\bibitem{CaCoMu}
R. D. Carlitz, J. C. Collins and A.H. Mueller,
\newblock {\it Phys. Lett.} {\bf B214}, 229 (1988).

\bibitem{AdlBard}
S. Adler and W. Bardeen,
\newblock {\it Phys. Rev.} {\bf 182}, 1517 (1969).

\bibitem{ELA}
E. Leader and M. Anselmino,
\newblock {\it Z. Phys.} {\bf C41}, 239 (1988).

\bibitem{Zijlstra}   
E. B. Zijlstra and W. L. van Neerven,
\newblock {\it Nucl. Phys.} {\bf B147}, 61 (1994).

\bibitem{AEL}
M. Anselmino, A. V. Efremov and E. Leader,
\newblock {\it Phys. Rep.} {\bf 261}, 1 (1995).

\bibitem{MulTer}
D. M\"{u}ller and O. V. Teryaev,
\newblock {\it Phys. Rev.} {\bf D56}, 2607 (1997).

\bibitem{Cheng}
Hai-Yang Cheng,
e-Print Archive:hep-ph/9712473.

\bibitem{preparation}
E. Leader, A. V. Sidorov and D. B. Stamenov, 
a paper in preparation.

\bibitem{GRV}
M. Gl\"{u}ck, E. Reya and A. Vogt,
\newblock {\it Z. Phys.} {\bf C67}, 433 (1995).

\bibitem{Parisi}
G. Parisi and N. Sourlas,
\newblock {\it Nucl. Phys.} {\bf B151}, 421 (1979);\\ 
I. S. Barker, C. B. Langensiepen and G. Shaw,
\newblock {\it Nucl. Phys.} {\bf B186}, 61 (1981).

\bibitem{Kriv}
V. G. Krivokhizhin et al.,
\newblock {\it Z. Phys.} {\bf C36}, 51 (1987);
{\it ibid} {\bf C48}, 347 (1990).

\bibitem{PDG}
Particle Data Group, L. Montanet et al.,
\newblock {\it Phys. Rev.} {\bf D50}, 1173 (1994);\\
F. E. Close and R. G. Roberts,
\newblock {\it Phys. Lett.} {\bf B313}, 165 (1993).

\bibitem{Jaffe}
R. L. Jaffe and A. Manohar,
\newblock {\it Nucl. Phys.} {\bf B337}, 509 (1990).

\bibitem{QCDsmallx}
R. D. Ball, S. Forte and G. Ridolfi,
\newblock {\it Nucl. Phys.} {\bf B444}, 287 (1995).

\bibitem{Ryskin}
J. Batrels, B. I. Ermolaev and M. G. Ryskin,
\newblock {\it Z. Phys.} {\bf C70}, 273 (1996);\\
{\bf C72}, 627 (1996).  

\end{thebibliography}
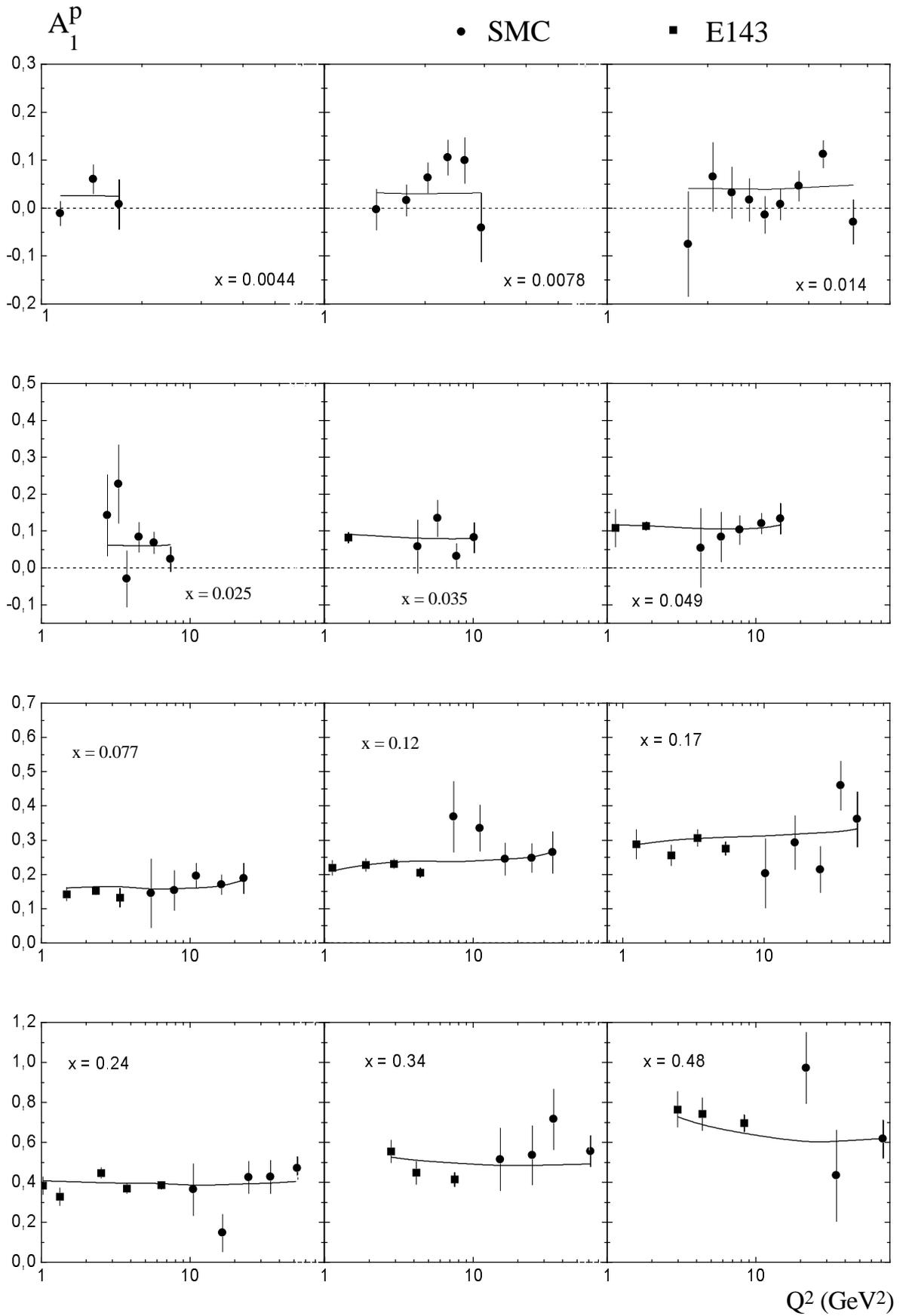
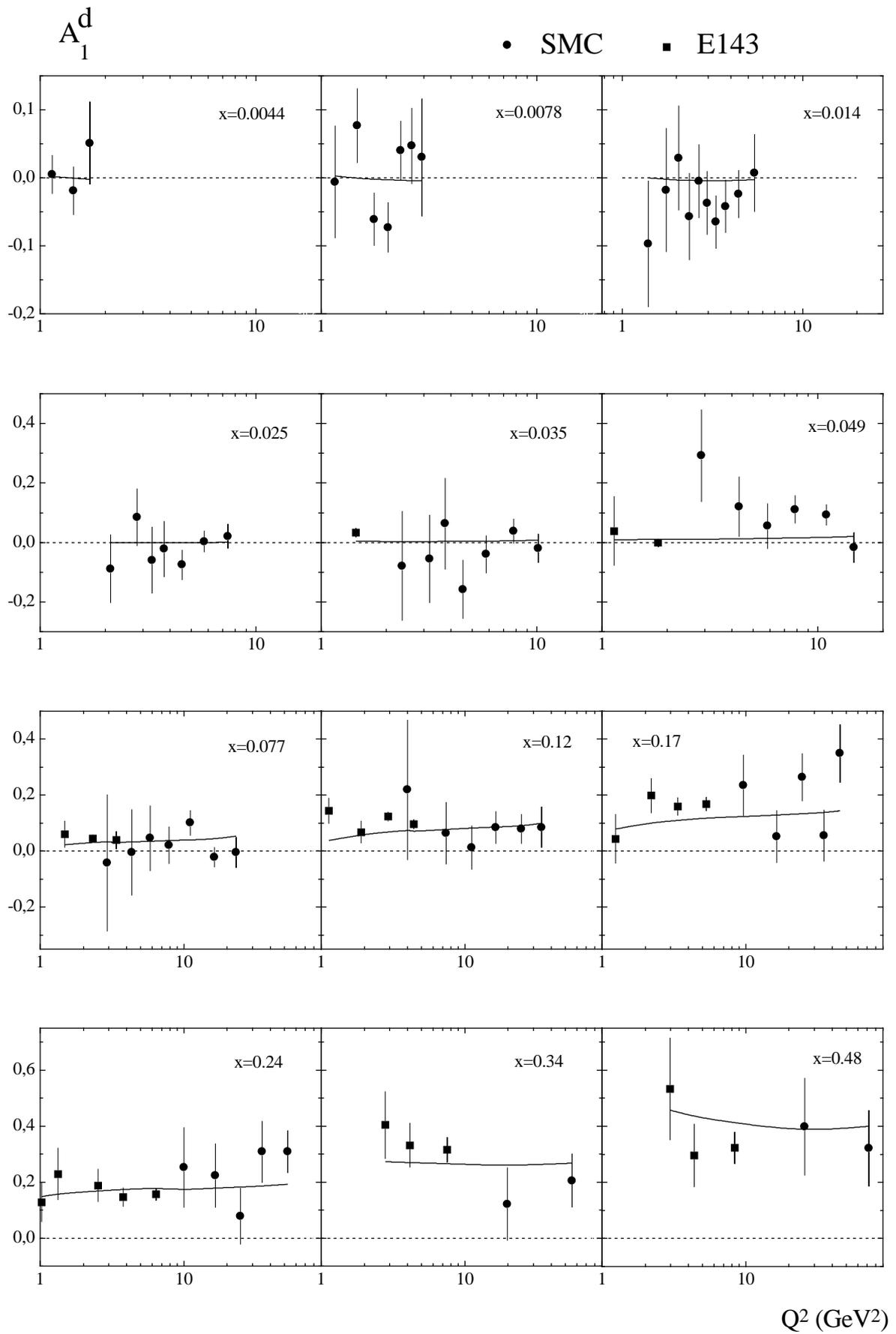
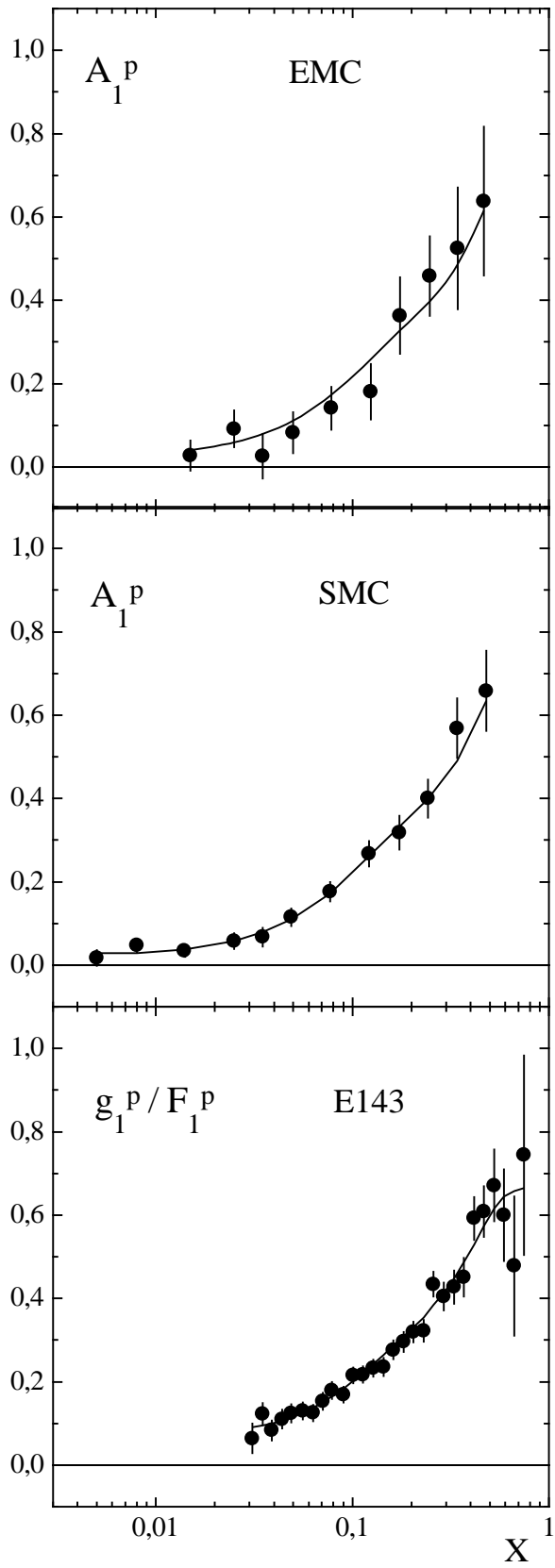
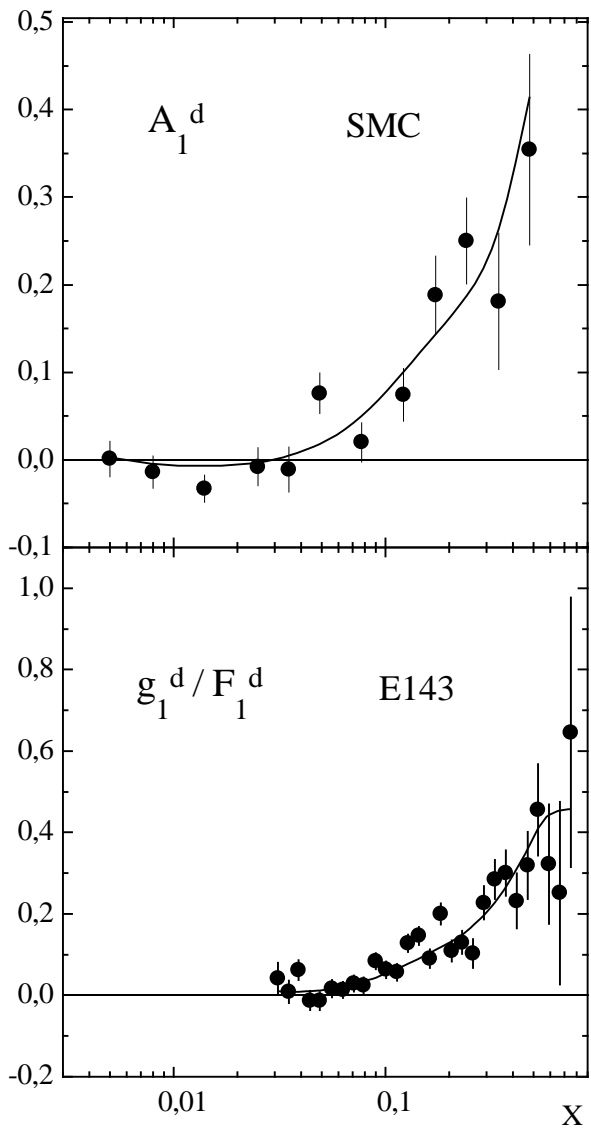
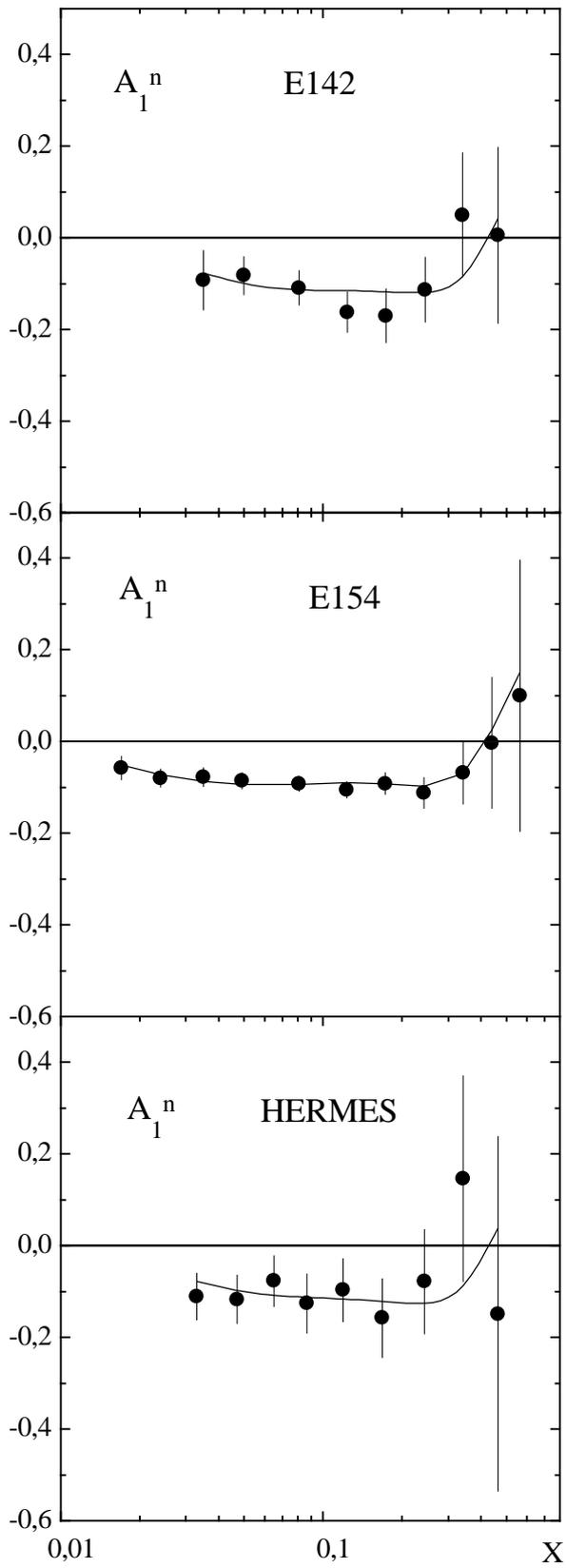
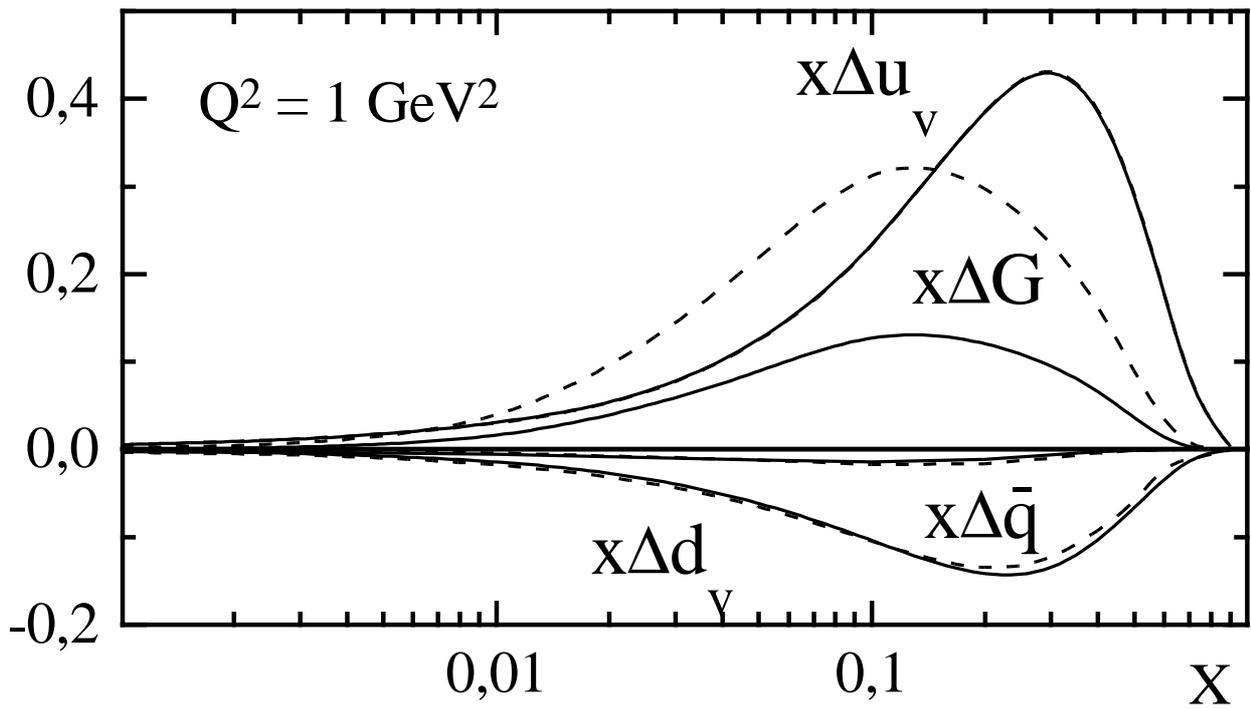
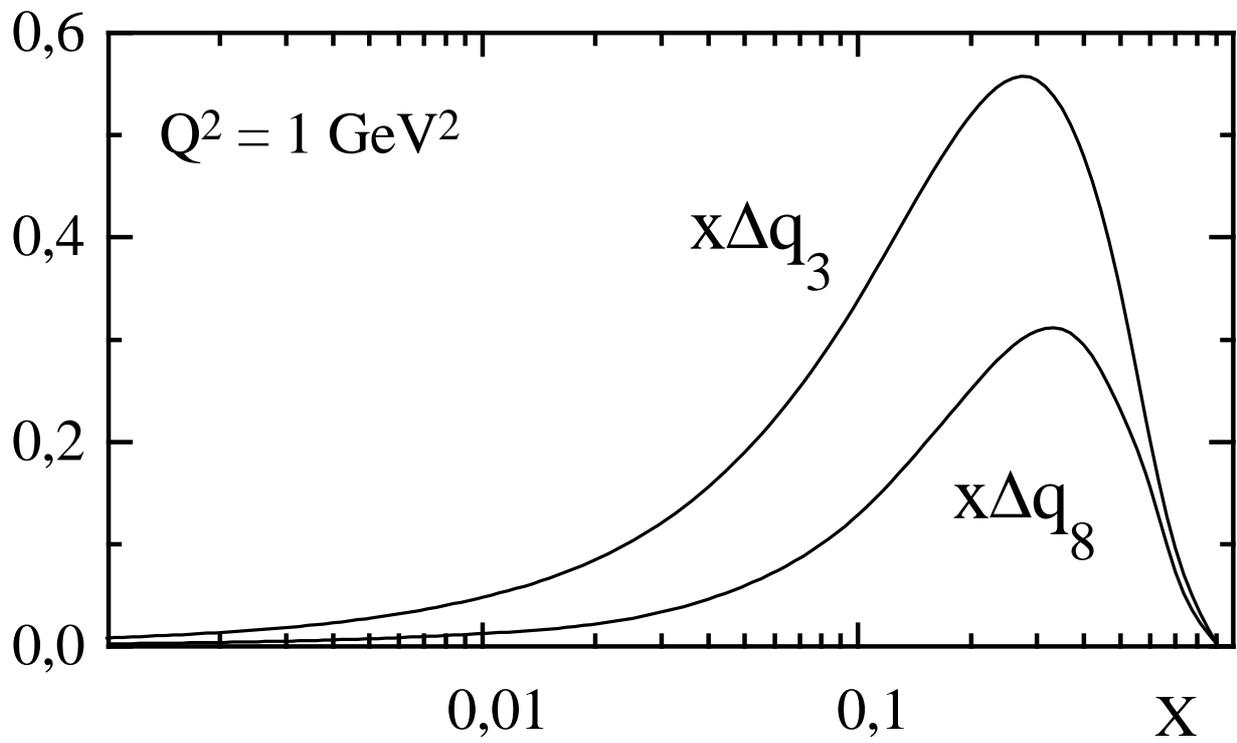
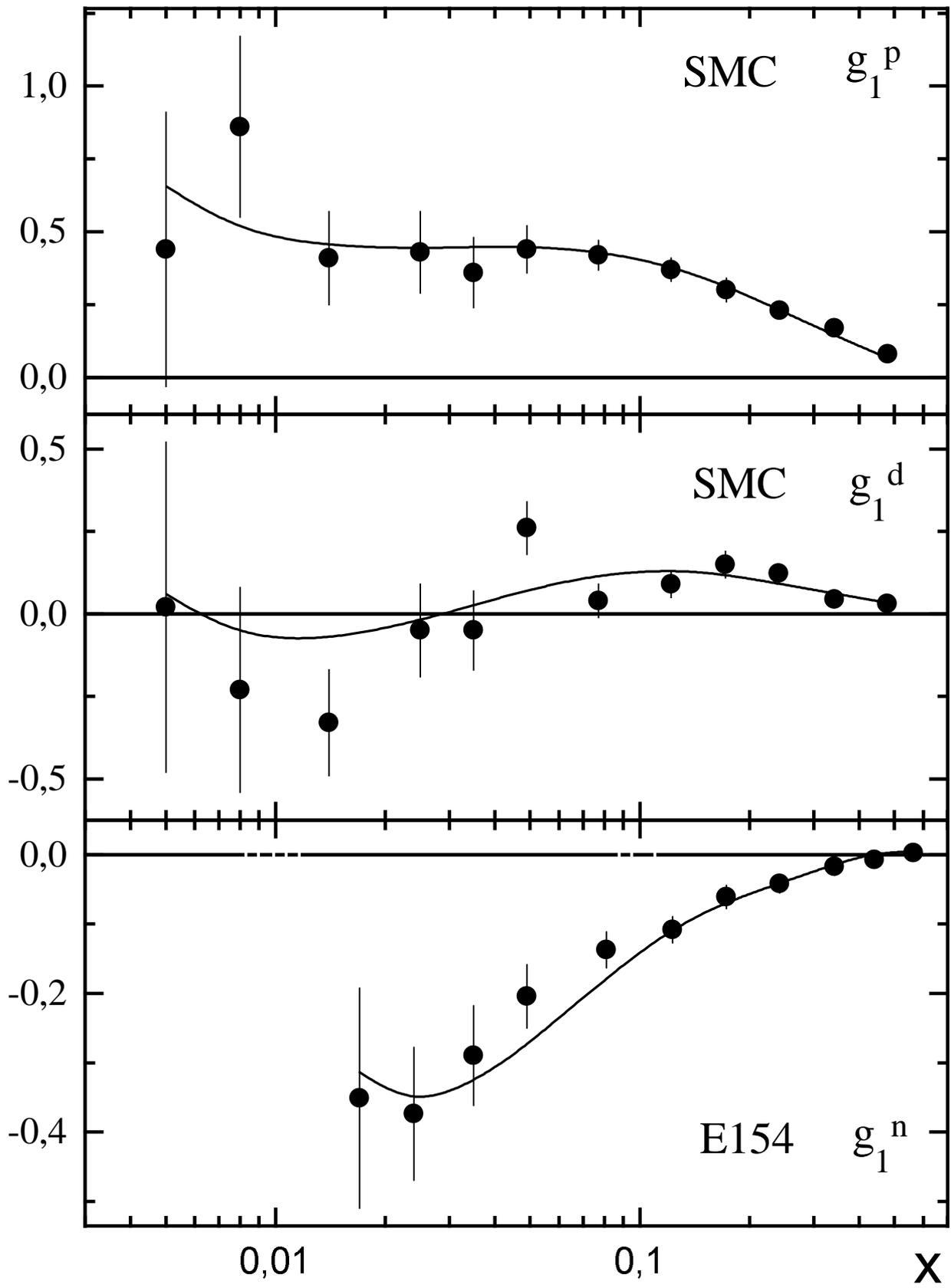
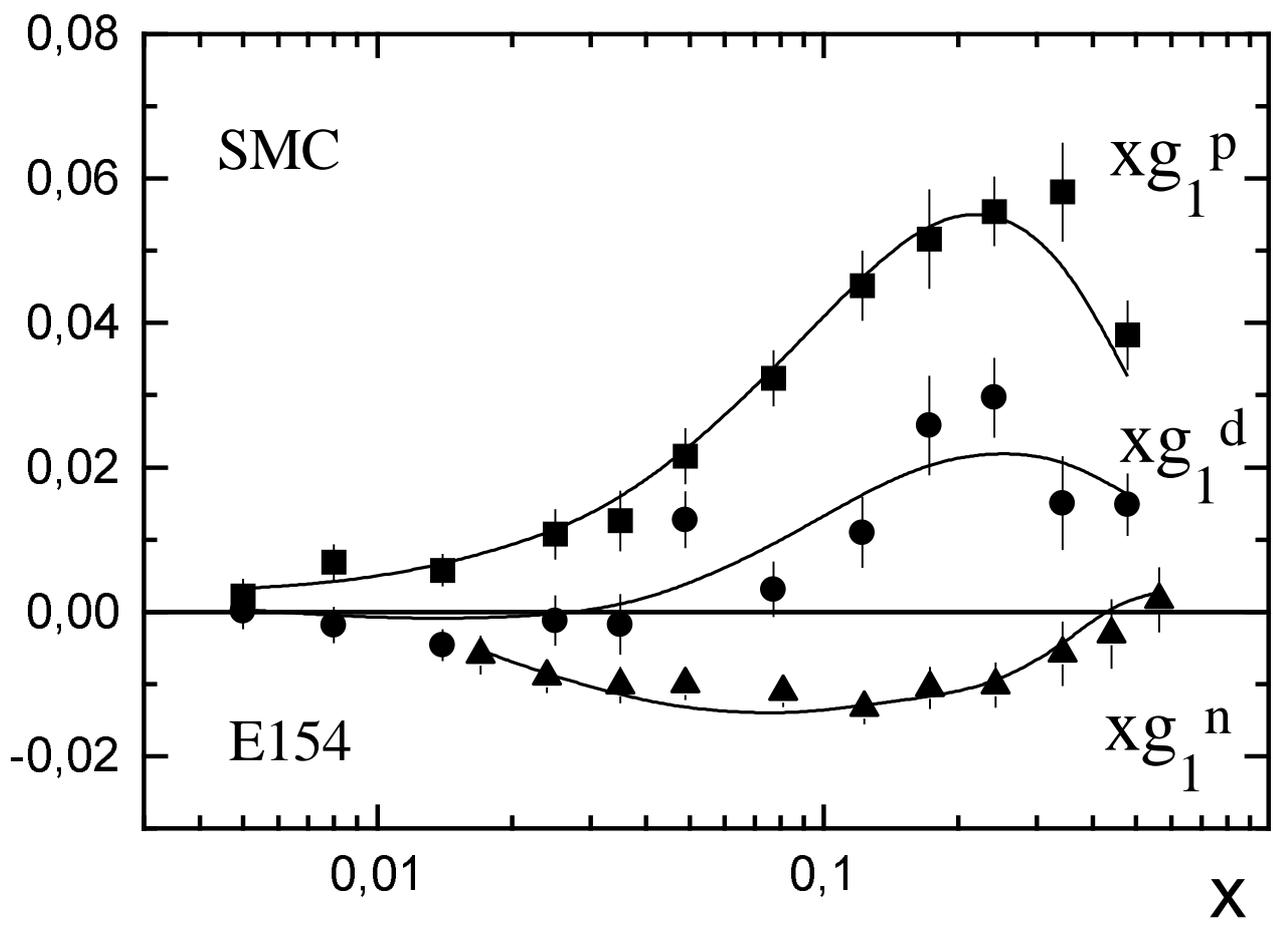
\end{document}